\documentclass[twocolumn]{article}
\usepackage{arxiv}
\usepackage[utf8]{inputenc}
\usepackage[T1]{fontenc}
\usepackage{cite}
\usepackage{amsmath,amssymb,amsfonts}
\usepackage{graphicx}
\usepackage{xcolor}
\usepackage{textcomp}
\usepackage{booktabs}
\usepackage{array}
\usepackage{tabularx}
\usepackage{microtype}
\usepackage{etoolbox}
\usepackage[hidelinks,hyperfootnotes=false]{hyperref}
\AtBeginEnvironment{thebibliography}{\clearpage\fontsize{8}{9}\selectfont\raggedright}
\makeatletter
\long\def\@makefntext#1{\noindent\raggedright#1}
\makeatother
\renewenvironment{abstract}
  {\par\small\noindent\textbf{Abstract}\par\smallskip\noindent\ignorespaces}
  {\par\medskip}

\newcommand{\authorrefmark}[1]{\ensuremath{^{\textbf{#1}}}}

\renewcommand{\shorttitle}{CTAG-FX}
\date{}

\title{CTAG-FX: Reinterpreting Synthesizer Parameter Spaces for
Expressive Tone-Shaping Audio FX Design}

\author{Geonung Jo\authorrefmark{1,*}, and Jongeun Choi\authorrefmark{2,*}\\[0.4em]
\normalfont\small\authorrefmark{1}Independent Researcher, Seoul, Republic of Korea\\
\normalfont\small\authorrefmark{2}School of Mechanical Engineering,\\
\normalfont\small Yonsei University, Seoul 03722, Republic of Korea
}

\hypersetup{
  pdftitle={CTAG-FX: Reinterpreting Synthesizer Parameter Spaces for Expressive Tone-Shaping Audio FX Design},
  pdfauthor={Geonung Jo, Jongeun Choi},
  pdfkeywords={Audio effects, synthesizer parameter mapping, text-guided audio processing, timbre processing}
}

\begin{document}
\twocolumn[\maketitle]
\thispagestyle{plain}
\begingroup
\renewcommand{\thefootnote}{}
\footnotetext{*These authors contributed equally to
this work. Corresponding author: Jongeun Choi
(email: jongeunchoi@yonsei.ac.kr).}
\endgroup

\begin{abstract}
Recent text-guided audio FX research has focused on translating natural-language descriptions into parameters or chain configurations within existing FX systems. However, comparatively little attention has been paid to how the internal control space of an individual FX processor can itself be constructed. To address this gap, we propose CTAG-FX, which functionally reinterprets the roles of 78 parameters in a text-conditioned synthesizer configuration as controls of a tone-shaping FX processor. Each synthesizer configuration defines a fixed tone-shaping processor that can be repeatedly applied to new audio inputs rather than producing only a one-off rendered output. The text-conditioned synthesizer configurations are generated using A\&R-CTAG, a retrieval-enhanced extension of CTAG. We evaluate CTAG-FX through signal-level analysis, together with a scrambled-mapping ablation in which parameter-to-control assignments are randomly permuted to assess the contribution of the proposed role assignment. Role-based mapping produces prompt-distinct spectral and nonlinear behavior, whereas scrambling reduces this prompt-specific differentiation and yields a more prompt-insensitive nonlinear profile. Overall, these results suggest that synthesizer parameter spaces can serve not only as sound-generation spaces but also as design resources for constructing new audio-FX control spaces. Audio samples are available at \url{https://taylor3527eg-hub.github.io/ctag-fx-platform-demo/}.
\end{abstract}


\section{Introduction}

Audio effects (FX) shape the timbre and character of musical signals. Desired sounds are often described using perceptual or metaphorical terms such as “warm,” “creamy,” or “crispy” \cite{zheng2016socialfx}, yet realizing these qualities requires selecting appropriate processing structures and adjusting low-level parameters. Text-guided audio FX seeks to translate such semantic intent directly into signal-processing controls. More broadly, controllable musical AI has emphasized responsiveness to performance input and explicit mechanisms for shaping desired musical attributes \cite{blanchard2024symbiotic,koo2025smitin}. In this context, the expressive capacity of text-guided FX depends not only on how language is mapped to parameters, but also on what processing and control space those parameters define.

Recent text-guided FX methods use Contrastive Language–Audio Pretraining (CLAP)-guided optimization, large language model (LLM)-based parameter prediction, black-box search, and iterative refinement to identify suitable FX settings from natural-language descriptions \cite{chu2025text2fx,doh2025llm2fx,ki2026fxsearcher,yu2026instructfx2fx}. Despite their different mechanisms, these approaches generally operate over predefined processor modules and their associated parameter or chain spaces. The central design problem is therefore typically how to navigate or control an available FX space, while comparatively little attention has been paid to how the internal control organization of an individual processor can itself be constructed.

Parametric synthesizers offer a different perspective on this problem. Their sound-generation architectures organize timbre formation through functionally differentiated components such as oscillators, envelopes, filters, mixers, and modulation structures. TorchSynth \cite{turian2021torchsynth} and SynthAX \cite{cherep2023synthax} enable exploration of high-dimensional synthesizer parameter spaces, while Creative Text-to-Audio Generation (CTAG) \cite{cherep2024ctag} uses text-conditioned optimization to instantiate sound characters as synthesizer configurations. These systems demonstrate that synthesizer parameter spaces can encode diverse sound characters through structured functional organization.

However, generator-side controls cannot be transferred directly to external-audio processing while retaining their original oscillator, envelope, and modulation functions. This leads to the central research question of this study: \textbf{Can the functional organization of a synthesizer parameter space be reinterpreted to construct a processor-level control space for tone-shaping audio FX?}

CTAG-FX addresses this question through role-based mapping, reinterpreting 78 synthesizer parameters as controls of an integrated tone-shaping processor rather than transferring their original synthesis functions. Each resulting configuration defines a fixed processor that can subsequently transform new external performance inputs.

Because CTAG-FX requires a text-conditioned synthesizer configuration as its input representation, we use Augmented and Retrieval-Based Creative Text-to-Audio Generation (A\&R-CTAG), a retrieval-enhanced extension of CTAG, to generate these synth-side configurations. A\&R-CTAG serves as the standardized synth-side representation generator; the processor-side design and evaluation addressed in this work concern CTAG-FX itself.

The main contributions of this study are as follows:

\begin{enumerate}
    \item We formulate text-guided FX design as a problem of processor control-space construction, using the functional organization of a synthesizer parameter space as a design resource rather than restricting semantic control to navigation within predefined FX spaces.
    
    \item We introduce a role-based reinterpretation of a 78-dimensional synthesizer parameter space into the controls of an integrated tone-shaping processor comprising drive, saturation, EQ, noise, mixing, and gain stages.
    
    \item Through signal-level analysis and a scrambled role-assignment ablation, we show that structured role mapping produces differentiated prompt-conditioned spectral and nonlinear behavior, while scrambled assignments reduce this differentiation.
\end{enumerate}

\section{Related Work}
\begin{figure}[tbp]
\centering

\includegraphics[width=1.00\columnwidth]{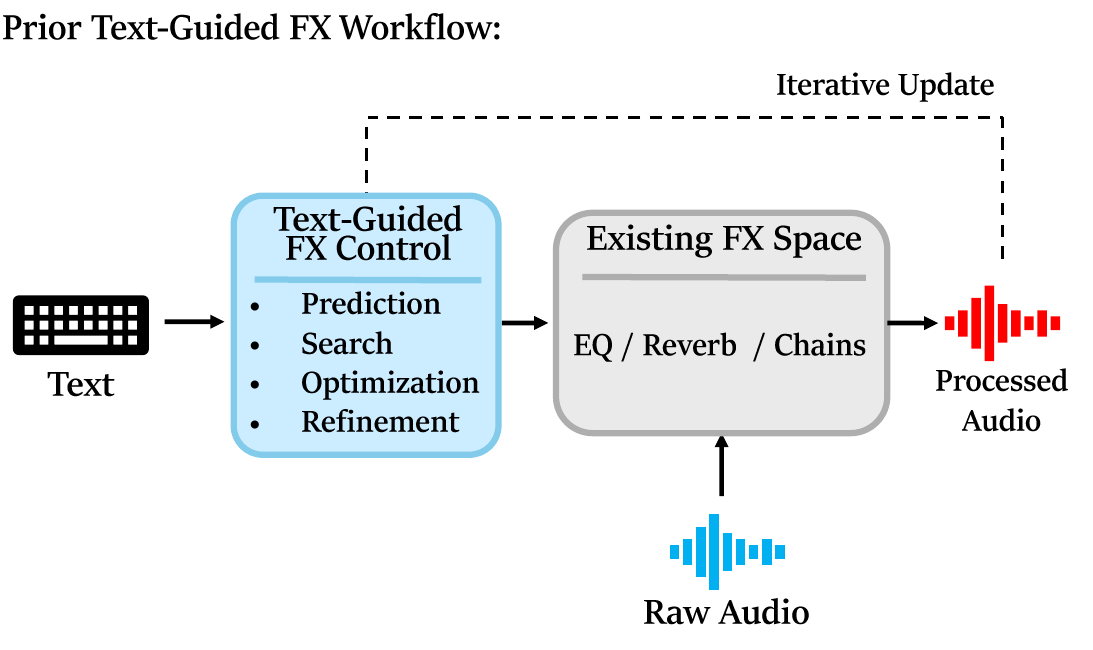}
\caption{Representative prior text-guided FX workflow.}
\label{fig:prior_workflow}

\vspace{0.4em}

\end{figure}

Recent text-guided audio FX research translates sonic intent expressed in natural language into FX parameters or chain configurations. Text2FX uses CLAP-guided differentiable optimization to identify FX configurations corresponding to textual descriptions \cite{chu2025text2fx}, while LLM2FX predicts FX parameters directly using a LLM \cite{doh2025llm2fx}. FxSearcher extends text-guided control to nondifferentiable processing modules through gradient-free black-box FX-chain search \cite{ki2026fxsearcher}. More recently, InstructFX2FX supports state-aware, multi-turn refinement of an existing FX chain \cite{yu2026instructfx2fx}. Fig.~\ref{fig:prior_workflow} illustrates this shared workflow.

Despite their different mechanisms, these approaches operate over predefined FX modules and their associated parameter or chain spaces. Their primary objective is to identify or update suitable configurations within an available processing space, rather than to construct the internal control organization of an individual processor.

Neural and differentiable audio-effect methods address adjacent tasks from a different perspective. Deep-learning black-box models have been used to emulate the input–output behavior of existing audio processors from measurements \cite{martinezramirez2020blackbox}. DeepAFx learns to control the parameters of existing stateful black-box effects for target audio-processing tasks \cite{martinezramirez2021deepafx}, while DeepAFx-ST predicts controls of differentiable audio effects to transfer production style from a reference recording \cite{steinmetz2022deepafxst}.  Therefore, these approaches focus on modeling, controlling, or transferring behavior within existing processing structures rather than constructing a new processor control organization.

Virtual-analog modeling similarly focuses on reproducing the behavior of existing analog processors in the digital domain. Recent work has combined Wave Digital Filters and neural approximations to model circuits containing multiple nonlinear elements in real time, including wah and fuzz processors \cite{giampiccolo2025virtualanalog}. By contrast, CTAG-FX constructs its processor control organization by functionally reinterpreting synthesizer parameter roles rather than emulating an existing processor or optimizing controls within a predefined FX space.

Therefore, CTAG-FX differs primarily in its design object: rather than modeling or controlling an existing processing structure, it uses synthesizer parameter organization to construct the internal control space of a tone-shaping processor (Table~\ref{tab:conceptual_comparison}). The comparison is conceptual because the systems address different objectives.


\begin{table*}[!t]
\centering
\caption{Conceptual comparison of prior text-guided FX systems and CTAG-FX.}
\label{tab:conceptual_comparison}
\footnotesize
\renewcommand{\arraystretch}{1.18}
\begin{tabular}{p{0.20\textwidth} p{0.38\textwidth} p{0.34\textwidth}}
\toprule
\textbf{Dimension} & \textbf{Prior text-guided FX systems} & \textbf{CTAG-FX} \\
\midrule
Strategy 
& Text $\rightarrow$ predefined FX configuration 
& Text $\rightarrow$ synth representation $\rightarrow$ FX processor design \\

Source of expressivity 
& Available FX modules and chains
& Synthesizer parameter space \\

Design space 
& Predefined FX parameter or chain space 
& Reinterpreted synth-to-FX control space \\

Main mechanism 
& Predict / optimize / search / refine 
& Reinterpret / map / design \\

Design object
& FX parameter or FX chain 
& Self-contained tone-shaping FX processor \\

Contribution
& Semantic control within predefined FX spaces
& Synth-derived FX control-space design \\
\bottomrule
\end{tabular}
\end{table*}

\section{Method}

\subsection{Overview}

\begin{figure*}[!t]
\centering
\includegraphics[width=0.98\textwidth]{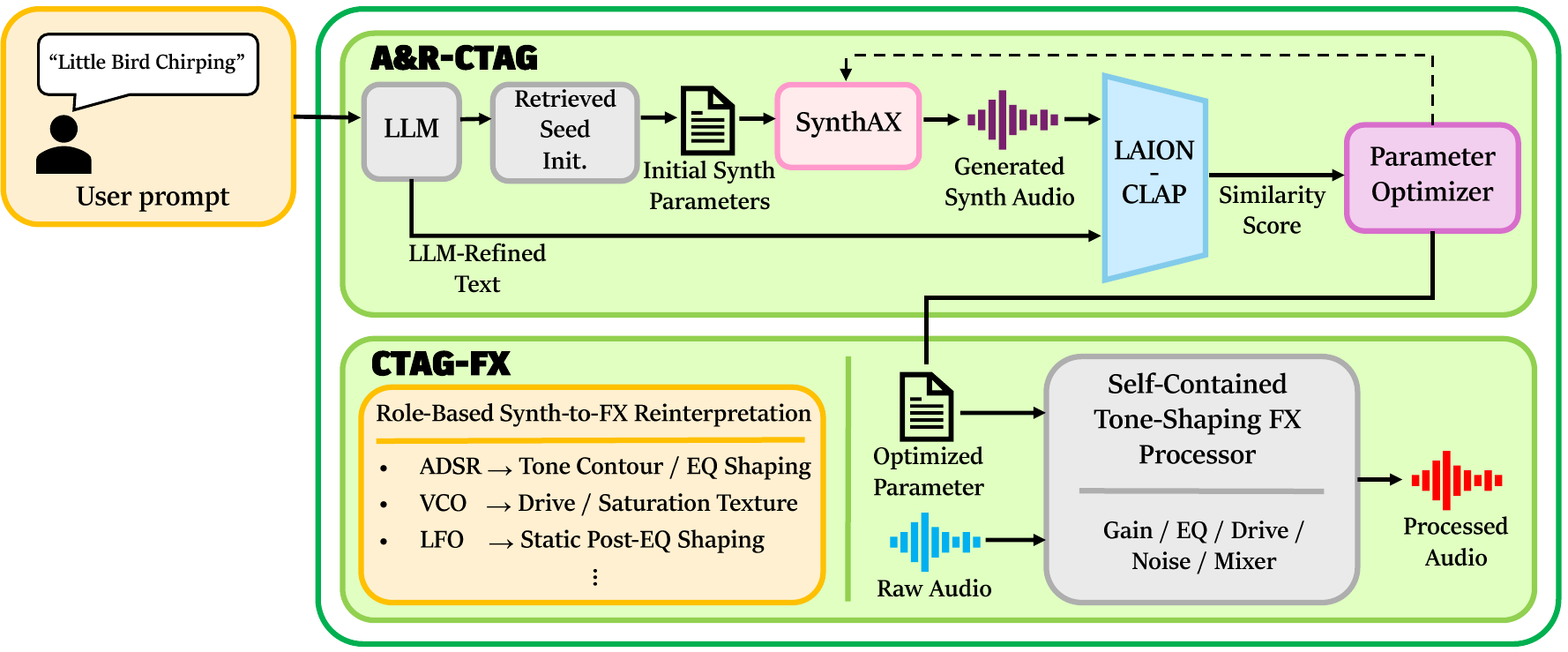}
\caption{Overall CTAG-FX workflow. A\&R-CTAG generates a text-conditioned synthesizer configuration, and CTAG-FX maps the functional roles of its parameter groups to controls in a tone-shaping processor for external audio.}
\label{fig:workflow}
\end{figure*}

CTAG-FX converts a text-conditioned synthesizer configuration generated by A\&R-CTAG into a fixed tone-shaping processor for external audio. As illustrated in Fig.~\ref{fig:workflow}, it reinterprets the roles of the synthesizer parameter groups as controls in an integrated chain comprising drive, EQ, noise, mixing, and input/output gain stages, rather than transferring the parameters directly. The resulting processor applies the configured tone-shaping behavior to external audio input.

\subsection{A\&R-CTAG}

\begin{figure}[tbp]
    \centering
    \includegraphics[width=\columnwidth]{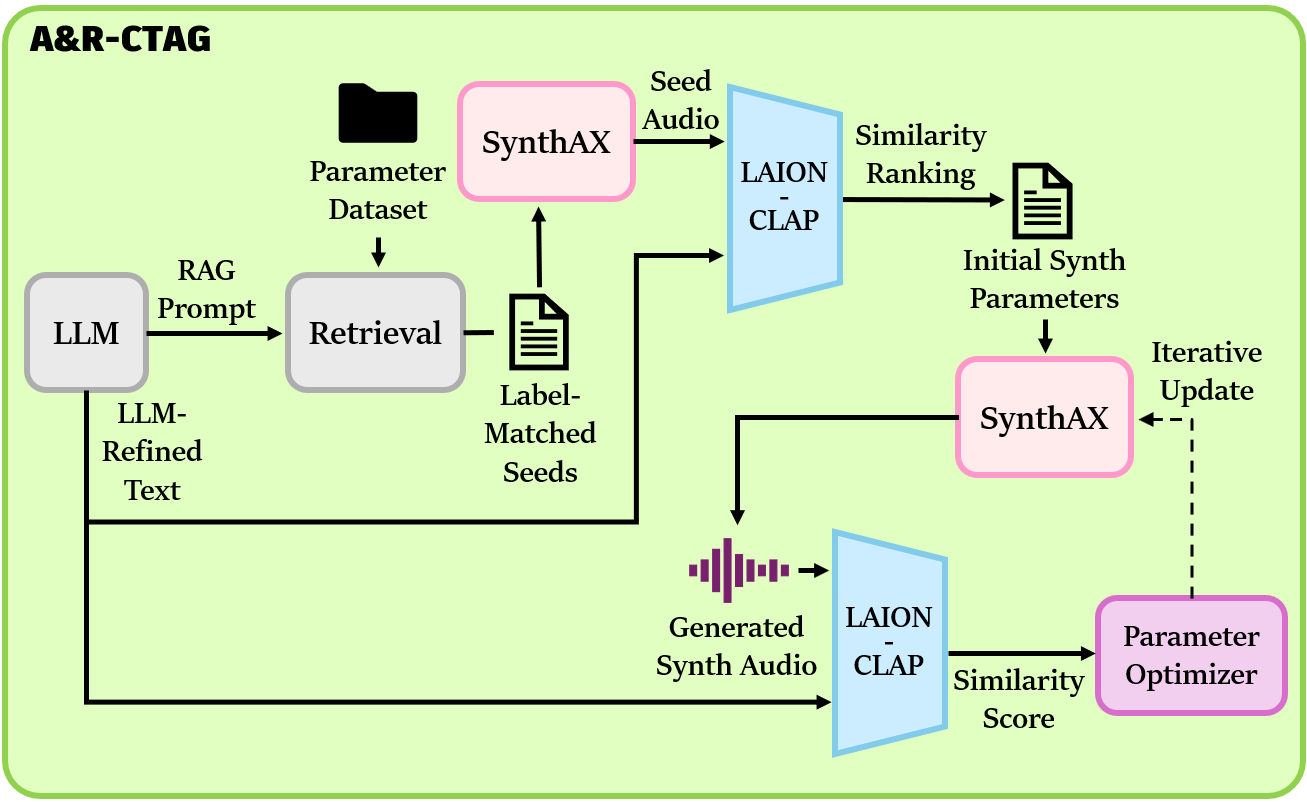}
    \caption{A\&R-CTAG pipeline for generating the synth-side character representation used by CTAG-FX.}
    \label{fig:A&R-CTAG}
\end{figure}

Fig.~\ref{fig:A&R-CTAG} illustrates the A\&R-CTAG pipeline, an extension of the original CTAG \cite{cherep2024ctag} that generates the synth-side character representation required by CTAG-FX. A\&R-CTAG retains the original CTAG optimization objective and learned evolution strategy (LES)-based optimizer \cite{lange2023evolution}, while augmenting its prompt processing and population initialization.

The LLM (GPT-4o) \cite{openai2024gpt4o} transforms the user prompt into a CLAP-aligned prompt for audio–text similarity evaluation and a retrieval prompt for semantic search. The retrieval model (all-MiniLM-L6-v2) \cite{sentenceTransformersMiniLM,reimers2019sentencebert} encodes the retrieval prompt and searches a dataset of 67 perceptually plausible sound-character configurations selected from outputs generated by the original CTAG \cite{cherep2024ctag}. The retrieved configurations are rendered using SynthAX \cite{cherep2023synthax} and re-ranked according to their LAION-CLAP \cite{wu2023laionclap} similarity scores with the aligned prompt. The highest-ranked configurations are then used as retrieval-based seeds.

The initial population combines equal proportions of retrieval-based and randomly initialized candidates, balancing semantic guidance with broader exploration. It is then optimized using the original CTAG procedure, and the highest-scoring configuration is passed to CTAG-FX as the synth-side character representation.

\subsection{Synth-to-FX Reinterpretation}

 The final A\&R-CTAG output \(\theta^{*}\) is a configuration in the 78-dimensional CTAG synthesizer parameter space \cite{cherep2024ctag}. Rather than treating it as a final synthesizer sound or assuming physical equivalence, CTAG-FX maps the character-forming roles of its parameter groups to processor-side controls. Three representative mappings follow.

\noindent\textbf{ADSR \(\rightarrow\) Pre-EQ / Tone Stack.}
In a synthesizer, attack, decay, sustain, and release (ADSR) parameters shape amplitude over time. Because CTAG-FX is a fixed-configuration processor, their character-forming tendencies are reinterpreted as static spectral-shaping controls rather than dynamic envelope generators.

\begin{figure}[tbp]
    \centering
    \includegraphics[width=0.8\columnwidth]{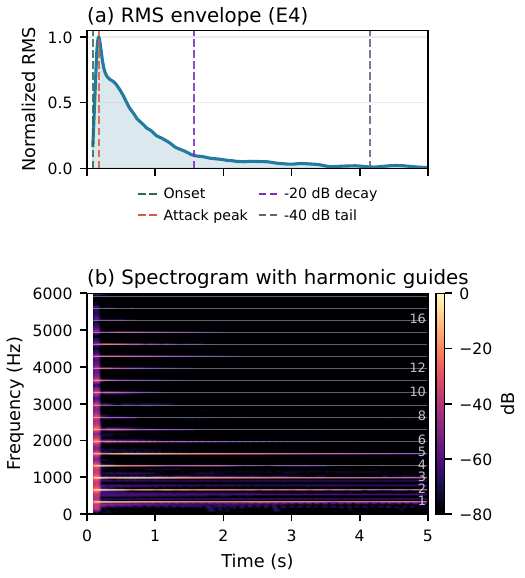}
    \caption{Temporal and spectral characteristics of a plucked E4 guitar note motivating the ADSR-to-EQ reinterpretation: (a) normalized RMS envelope with temporal markers, (b) spectrogram with harmonic guides.}
    \label{fig:ADSR-Harmonics}
\end{figure}

Temporal-envelope and spectral cues jointly contribute to perceived timbre \cite{caclin2005timbre,kazazis2021descriptors,woodhouse2004guitar}. As shown in Fig.~\ref{fig:ADSR-Harmonics}, a plucked E4 note exhibits an ADSR-like RMS contour while its harmonic balance evolves over time. This temporal–spectral relationship motivates a static reinterpretation in which the primary ADSR group controls Pre-EQ bandwidth and spectral balance, while the secondary group controls the Tone Stack’s bass, midrange, treble, and presence.

\begin{figure}[tbp]
\centering
    \includegraphics[width=0.8\columnwidth]{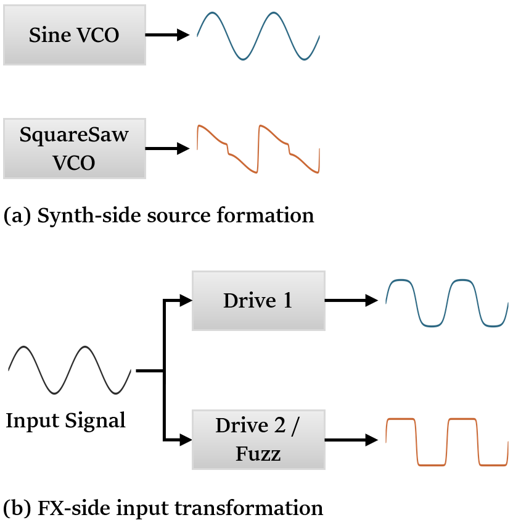}
    \caption{Conceptual VCO-to-drive reinterpretation. (a) Sine and SquareSaw VCOs generate comparatively smooth and harmonically rich sources, respectively. (b) Drive~1 and Drive~2/Fuzz apply analogous smooth and aggressive nonlinear coloration to a common input. Colors indicate corresponding roles; waveforms are illustrative, not equivalent.}
    \label{fig:vco_drive_reinterpretation}
\end{figure}

\noindent\textbf{VCO \(\rightarrow\) Drive / Fuzz.}
Voltage-controlled oscillators (VCO) serve as primary timbral sources in synthesizers through waveforms with distinct harmonic structures. Because CTAG-FX processes external audio, this source-forming role is reinterpreted as nonlinear drive rather than oscillator generation. As illustrated in Fig.~\ref{fig:vco_drive_reinterpretation}, the Sine VCO group is mapped to Drive~1 for comparatively smooth nonlinear coloration, while the SquareSaw VCO group is mapped to Drive~2/Fuzz for more aggressive, harmonically rich shaping. This preserves the contrast between their character-forming roles without implying oscillator-function or waveform equivalence.

\noindent\textbf{Mod Matrix.}
Whereas the ADSR and VCO mappings preserve the character-forming roles of individual module groups, the Mod Matrix preserves their influence relationships. The original TorchSynth topology \cite{turian2021torchsynth} comprises four sources—two ADSR envelopes and two low-frequency oscillators (LFO)—and five targets controlling the pitch and amplitude of the two VCOs and the noise amplitude. CTAG-FX retains this four-source–five-target structure as static cross-axis control: the source groups are summarized as Pre-EQ, Tone Stack, Post-EQ High, and Post-EQ Low character axes and routed to Drive 1 gain, Drive 2 gain, Pre-EQ high-pass filter (HPF), Post-EQ High gain, and noise level.

Let the normalized Mod Matrix parameters be
\(W_0 \in [0,1]^{5 \times 4}\). It is converted into a bipolar routing-weight matrix as
\begin{equation}
    W
    =
    2\operatorname{clip}(W_0,0,1)-1,
    \qquad
    W \in [-1,1]^{5 \times 4}.
    \label{eq4}
\end{equation}

Each source group is reduced to a bipolar scalar summary reflecting its reinterpreted FX role. The four summaries are arranged as
\[
m =
\left[
m_{\mathrm{Pre}},
m_{\mathrm{Tone}},
m_{\mathrm{High}},
m_{\mathrm{Low}}
\right]^{\top}
\in [-1,1]^4.
\]

The five-dimensional target influence vector is then calculated as
\begin{equation}
    t = \operatorname{clip}(Wm,-1,1),
    \qquad
    t \in [-1,1]^5.
    \label{eq5}
\end{equation}

Each element of \(t\) is applied as a bounded offset to the corresponding Drive, EQ, or Noise control. The Mod Matrix therefore functions as a static cross-axis routing layer rather than an audio-processing block.

The same role-based principle is applied to the remaining parameter groups. Table~\ref{tab:synth_to_fx_mapping} summarizes their original synth-side roles, functional reinterpretations, and corresponding FX controls. Collectively, these mappings convert the 78-dimensional synthesizer configuration into controls for a single tone-shaping processor while preserving module-group distinctions and Mod Matrix routing relationships.

\subsection{FX Processor Design}

\begin{figure}[tbp]
    \centering
    \includegraphics[width=\columnwidth]{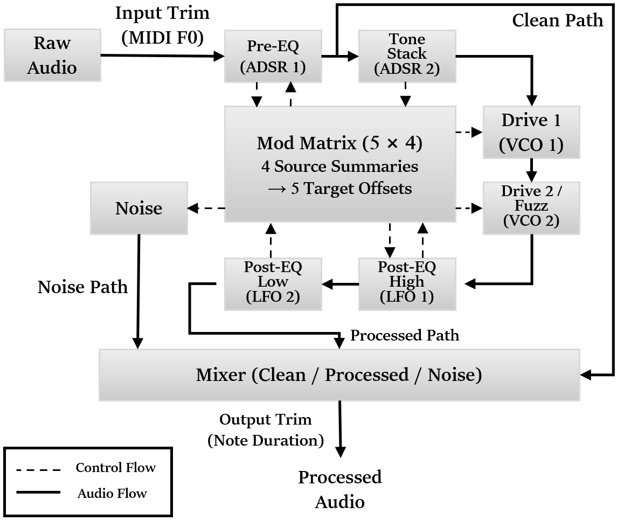}
    \caption{Simplified CTAG-FX signal flow with synth-derived control routing. Parentheses identify the corresponding synthesizer groups.}
    \label{fig:CTAG-FX Processor}
\end{figure}

Fig.~\ref{fig:CTAG-FX Processor} illustrates the signal-flow architecture instantiated by the reinterpreted control groups. The stages are arranged so that the spectral conditions entering the nonlinear blocks and the subsequent shaping of generated harmonics can be controlled separately.

The input first passes through Input Trim and Pre-EQ. Input Trim sets the level delivered to the processor, while Pre-EQ determines the bandwidth and spectral balance entering the nonlinear stages. The signal is then divided into clean and main processing paths. In the main path, the Tone Stack precedes the serial Drive~1 and Drive~2 stages and shapes the spectrum before harmonic generation. Drive~1 provides primary saturation and harmonic coloration, whereas Drive~2 adds more pronounced distortion and fuzz-like texture. Post-EQ High and Post-EQ Low then shape the generated harmonics and determine the final brightness and body without altering the input conditions of the drive stages.

The Mod Matrix operates outside the audio path, applying the four-to-five static cross-axis routing defined above to the Pre-EQ, Drive, Post-EQ, and Noise controls.

The Mixer combines the clean, processed, and independent noise paths, followed by Output Trim and an auxiliary RMS/peak stabilization stage informed by ITU-R BS.1770 \cite{iturbs1770} without implementing its complete loudness or true-peak algorithm. The stabilized signal is returned as the final processed output.

Because each configuration defines a fixed feed-forward processor, its input--output behavior can also be captured using Neural Amp Modeler (NAM) \cite{atkinson2023nam}. For a separate NAM capture test, eight CTAG-FX configurations were used: one configuration from each of the four tone-shaping prompts evaluated in Sec.~4.1 and four additional configurations generated from broader sound-character prompts. These configurations were captured at 48~kHz on TONE3000 using the A2 architecture and 200 training epochs, yielding final error-to-signal ratio (ESR) values of 0.0003--0.0815 (median 0.0140).

\begin{table}[tbp]
\centering
\caption{Summary of synthesizer groups, functional reinterpretations, and FX controls in CTAG-FX.}
\label{tab:synth_to_fx_mapping}
\footnotesize
\setlength{\tabcolsep}{3pt}
\renewcommand{\arraystretch}{1.10}

\begin{tabularx}{\columnwidth}{@{}
>{\raggedright\arraybackslash}p{0.19\columnwidth}
>{\raggedright\arraybackslash}p{0.48\columnwidth}
>{\raggedright\arraybackslash}X
@{}}
\toprule
\textbf{Synth Group}
& \textbf{Functional Reinterpretation}
& \textbf{FX Control} \\
\midrule

MIDI \(f_0\)
& Pitch reference\newline
  \(\rightarrow\) input-response reference
& Input Trim \\

Duration
& Event length\newline
  \(\rightarrow\) output-scale reference
& Output Trim \\

ADSR1
& Temporal envelope\newline
  \(\rightarrow\) spectral contour
& Pre-EQ \\

ADSR2
& Secondary envelope\newline
  \(\rightarrow\) complementary tone contour
& Tone Stack \\

VCO1/2
& Waveform source\newline
  \(\rightarrow\) nonlinear harmonic character
& Drive 1/2 \\

Mixer
& Source balance\newline
  \(\rightarrow\) processing-path balance
& Clean / Processed / Noise Mixer\\

LFO1/2
& Time-varying modulation\newline
  \(\rightarrow\) static tonal shape
& Post-EQ High/Low \\

Mod Matrix
& Modulation routing\newline
  \(\rightarrow\) static cross-axis influence
& Five target controls \\

\bottomrule
\end{tabularx}
\end{table}

\subsection{Evaluation Setup}

The evaluation addressed processor-side tone shaping and synth-side representation generation separately:

\begin{enumerate}
    \item \textbf{Tone-Shaping FX Operation Test:}
    To isolate the processor-side functionality of CTAG-FX, this test bypassed A\&R-CTAG. Following the direct language-to-parameter paradigm of LLM2FX~\cite{doh2025llm2fx}, GPT-4o generated CTAG-FX configurations directly from each tone prompt and the defined parameter schema.
    The LLM was used only to instantiate prompt-conditioned configurations; the evaluation examined their spectral, harmonic, and dynamic behavior across prompts. A scrambled-mapping ablation assessed the contribution of the proposed parameter-to-control assignment.

    \item \textbf{Synth-Side Representation Generator Comparison:}
    CTAG and A\&R-CTAG were compared using LAION-CLAP similarity under the same optimization budget. This comparison evaluated synth-side representation generation separately from the processor-side signal analysis.
\end{enumerate}

All experiments used 48-kHz audio on an AMD Ryzen~5 5600H CPU, 32~GB RAM, and an NVIDIA RTX~3060 Laptop GPU.

\section{Evaluation}

\subsection{Tone-Shaping FX Operation Test}

\begin{figure*}[!t]
\centering
\includegraphics[width=1.00\textwidth]{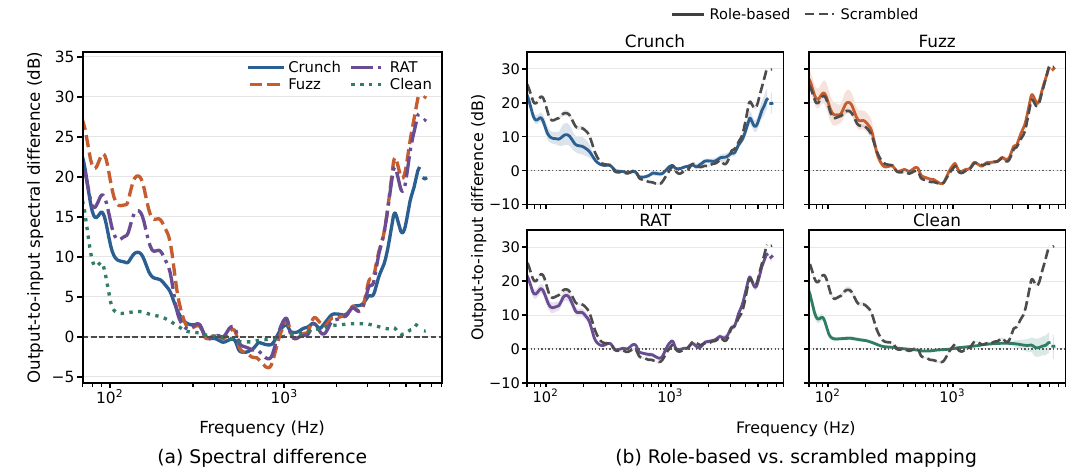}
    \caption{Spectral characterization and mapping ablation for four tone prompts. (a) Mean output-to-input spectral difference across ten configurations. (b) Role-based versus scrambled-mapping responses. Shading indicates 95\% confidence intervals.}
    \label{fig:tone_shaping_signatures}
\end{figure*}

Four tone prompts covering clean, crunch, fuzz, and distortion characters were evaluated: Chewy Amp Crunch, Germanium Fuzz, RAT-style Distortion, and Soft Creamy Clean. For each prompt, ten FX configurations were independently generated and applied to the same held-out 10-s clean guitar performance. These configurations were treated as repeated samples of prompt-conditioned processor behavior.

To examine whether the four prompts produced distinguishable tone-shaping characteristics, the resulting outputs were analyzed across spectral, harmonic, and dynamic dimensions. Output-to-input spectral differences were computed over the full performance after RMS matching each output to the dry input, excluding bins more than 60~dB below the dry spectral peak. This analysis characterized the frequency regions emphasized or attenuated by each prompt-conditioned processor. Furthermore, to test whether the observed prompt-dependent behavior depended on the proposed role assignment, we evaluated scrambled mappings. For each of the ten configurations per prompt, the normalized 78-dimensional parameter vector was retained while its parameter-to-control assignment was randomly permuted using 30 fixed permutations shared across all prompts and configurations. Each scrambled output was processed and analyzed using the same input and pipeline as the role-based condition. The 30 permutation results were averaged within each configuration, preserving the original configuration as the statistical unit (\(N=10\) per prompt). Nonlinear behavior was quantified using relative harmonic-energy change, H6–H10 share, and active-region crest-factor reduction, with the active region defined from the dry signal by a 20-ms RMS envelope at 30 dB below its maximum. Prompt-level means and 95\% confidence intervals (CI) were computed across the ten independently generated configurations.

As shown in Fig.~\ref{fig:tone_shaping_signatures}(a), the four prompt conditions produced distinct spectral responses, although the three drive-oriented prompts shared broadly similar broadband contours. The role-based versus scrambled comparison in Fig.~\ref{fig:tone_shaping_signatures}(b) revealed a prompt-dependent effect of role assignment. Scrambling produced substantial spectral changes for Clean and Crunch, more localized changes for RAT, and comparatively small changes for Fuzz. Over 70 Hz–8 kHz, the mean absolute differences between the role-based and scrambled mean responses were 6.39 dB for Clean, 2.75 dB for Crunch, 1.48 dB for RAT, and 0.80 dB for Fuzz. The weak separation for Fuzz suggests that strong nonlinear saturation may make long-term spectral behavior less sensitive to parameter-role reassignment, or that this spectral measure does not fully expose differences within that processing regime. Of the 1,200 scrambled renders, 28 produced non-finite outputs due to numerical instability under particular permuted assignments and were excluded. The remaining 26–30 permutations per original setting were averaged first, preserving the ten independently generated settings as the statistical units (\(N=10\) per prompt).

\begin{figure}[tbp]
    \centering
    \includegraphics[width=0.9\columnwidth]{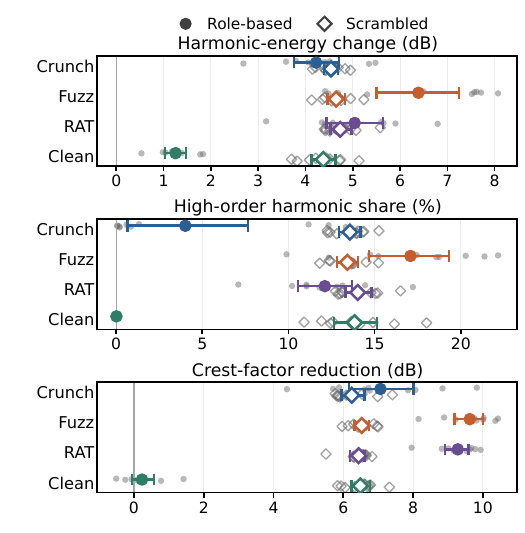}
    \caption{Nonlinear descriptors under role-based and scrambled mappings. Markers and error bars show configuration-level means and 95\% confidence intervals (\(N=10\) per prompt).}
    \label{fig:nonlinear descriptors}
\end{figure}

Fig. 8 provides a clearer view of this effect in the nonlinear descriptor space. Rather than retaining the prompt-specific profiles observed under role-based mapping, the scrambled mappings exhibited a largely prompt-insensitive nonlinear profile, and their descriptor centroids occupied a relatively narrow common region. Under role-based mapping, Fuzz showed the largest harmonic-energy increase and high-order harmonic share, Fuzz and RAT exhibited the strongest crest-factor reductions, and Clean remained closest to the dry input. These distinctions were substantially reduced after scrambling.

\subsection{Synth-Side Representation Generator Comparison}

CTAG and A\&R-CTAG were compared on ten prompts under the same 300-iteration budget; A\&R-CTAG used four retrieved seeds in a population of 100 candidates equally divided between retrieval-based and random initialization. As shown in Table~\ref{tab:ar_ctag_comparison}, A\&R-CTAG achieved higher and less variable final LAION-CLAP similarity. Because the initial-to-final gains were similar, the improvement was primarily attributable to retrieval-enhanced initialization rather than greater optimization progress. Overall, these results support the use of A\&R-CTAG as a more reliable synth-side representation generator for CTAG-FX.

\begin{table}[tbp]
\caption{LAION-CLAP similarity results for CTAG and A\&R-CTAG across ten prompts under a 300-iteration budget. Mean scores and final-score SD are computed across prompts; Change is relative to CTAG.}
\label{tab:ar_ctag_comparison}
\centering
\footnotesize
\setlength{\tabcolsep}{4pt}
\renewcommand{\arraystretch}{1.08}
\begin{tabularx}{\columnwidth}{@{}Xccc@{}}
\toprule
\textbf{Metric}
& \textbf{CTAG}
& \textbf{A\&R-CTAG}
& \textbf{Change} \\
\midrule
Initial score mean
& 0.1449 & 0.2981 & $+105.7\%$ \\

Final score mean
& 0.6166 & 0.7545 & $+22.4\%$ \\

Final score SD
& 0.1246 & 0.0595 & $-52.2\%$ \\

Score change mean
& 0.4718 & 0.4564 & $-3.3\%$ \\

\bottomrule
\end{tabularx}
\end{table}

\section{Limitations and Future Work}

 This study has three main limitations. First, the proposed synth-to-FX mapping is a heuristic functional reinterpretation rather than a unique, optimal, physically equivalent, or perceptually calibrated correspondence. The scrambled-mapping ablation supports the importance of structured role assignment, but does not establish the chosen assignment as optimal. Further calibration may therefore be needed, and synth-parameter magnitudes should not be interpreted as quantitatively accurate or perceptually linear FX control intensities.

Second, CTAG-FX is currently limited to fixed-configuration tone shaping using drive, saturation, EQ, noise, and static mixing, excluding explicit delay, reverb, and time-varying modulation. Robustness to extreme control combinations also requires further refinement, as indicated by the non-finite outputs observed under a small subset of scrambled assignments.

Third, the processor-side evaluation was limited to four tone prompts, ten configurations per prompt, and a single 10-s guitar performance. It examined configurations generated directly from tone prompts, while A\&R-CTAG was assessed separately using LAION-CLAP similarity. The reported signal-level differences and embedding similarities do not establish generalization across instruments and performances or perceptual correspondence between text descriptions, synthesized sounds, and processed audio.

An IRB-approved user study is planned to assess the perceptual plausibility of the generated tone-shaping effects and the correspondence between text descriptions, synthesized sounds, and processed audio. The results will be incorporated into a future version of this work.

\section{Conclusion}

This work presented CTAG-FX, which extends text-guided audio FX from selecting configurations within predefined processing spaces toward constructing processor control spaces through functional reinterpretation of synthesizer parameter roles. Role-based mapping produced differentiated prompt-conditioned spectral and nonlinear behavior, whereas scrambling reduced this differentiation, supporting the importance of structured role assignment without implying that the proposed mapping is unique or optimal.

Because each configuration defines a reusable fixed processor that can also be captured using NAM for real-time use, the results suggest that synthesizer parameter spaces can serve not only as sound-generation spaces but also as design resources for constructing new audio-effect control spaces.

\end{document}